\documentclass[sigconf, nonacm]{acmart}

\usepackage{xcolor}
\usepackage{enumitem}
\usepackage{amsmath}

\usepackage{subfiles}
\usepackage{blindtext} 
\usepackage{listings}
\usepackage{array}
\usepackage{xspace}
\usepackage[ruled,vlined,linesnumbered]{algorithm2e}

\usepackage{graphicx}
\usepackage{balance}
\graphicspath{{images/}{../images/}}

\newcommand{\jeg}[1]{}

\newcommand\vldbdoi{10.14778/3436905.3436925}
\newcommand\vldbpages{XXX-XXX}
\newcommand\vldbvolume{14}
\newcommand\vldbissue{4}
\newcommand\vldbyear{2021}
\newcommand\vldbauthors{\authors}
\newcommand\vldbtitle{\shorttitle} 
\newcommand\vldbavailabilityurl{https://github.com/ucbrise/flor}
\newcommand\vldbpagestyle{empty}

\newcommand{\record}{record\xspace}
\newcommand{\replay}{replay\xspace}

\newcommand{\smallitem}[1]{\vspace{0.3em}\noindent\textbf{#1}}
\newcommand{\botitem}[1]{\noindent\textbf{#1}}

\newcommand{\skipblock}{SkipBlock\xspace}
\newcommand{\flor}{\texttt{flor}\xspace}

\definecolor{brewerpurple}{HTML}{AF4EA3}
\definecolor{brewerblue}{HTML}{377EB8}
\definecolor{brewergreen}{HTML}{4DAF4A}
\definecolor{revisionscolor}{rgb}{0.0, 0.0, 0.0}

\lstdefinestyle{customc}{
    emph={checkpoint_resume,skipblock,tensorboard},emphstyle=\bfseries\color{brewerpurple},morekeywords={yield},
    emph={[2]args},emphstyle={[2]\bfseries\color{brewerblue}},
    language=Python,numbers=left,keywordstyle=\bfseries\color{green!40!black},numberstyle=\color{gray}
}

\begin{document}


\title{Hindsight Logging for Model Training}

\author{Rolando Garcia}
\affiliation{%
  \institution{UC Berkeley}
}
\email{rogarcia@berkeley.edu}

\author{Eric Liu}
\affiliation{%
  \institution{UC Berkeley}
}
\email{rickchang@berkeley.edu}

\author{Vikram Sreekanti}
\affiliation{%
  \institution{UC Berkeley}
}
\email{vikrams@berkeley.edu}

\author{Bobby Yan}
\affiliation{%
  \institution{UC Berkeley}
}
\email{bobbyy@berkeley.edu}

\author{Anusha Dandamudi}
\affiliation{%
  \institution{UC Berkeley}
}
\email{adandamudi@berkeley.edu}

\author{Joseph E. Gonzalez}
\affiliation{%
  \institution{UC Berkeley}
}
\email{jegonzal@berkeley.edu}

\author{Joseph M. Hellerstein}
\affiliation{%
  \institution{UC Berkeley}
}
\email{hellerstein@berkeley.edu}

\author{Koushik Sen}
\affiliation{%
  \institution{UC Berkeley}
}
\email{ksen@berkeley.edu}

\begin{abstract}

    In modern Machine Learning, model training is an iterative, experimental process 
    that can consume enormous computation resources and developer time.
    To aid in that process, experienced model developers log and visualize
    program variables during training runs. Exhaustive logging of all 
    variables is infeasible, so developers are left to choose between slowing 
    down training via extensive \emph{conservative} logging, or letting training run
    fast via minimalist \emph{optimistic} logging that may omit 
    key information. As a compromise, optimistic logging can be accompanied 
    by program checkpoints; this allows developers to add log statements 
    post-hoc, and ``replay'' desired log statements from checkpoint---a 
    process we refer to as \emph{hindsight} logging. Unfortunately, 
    hindsight logging raises tricky problems in data management and software engineering. 
    Done poorly, hindsight logging can waste resources and
    generate technical debt embodied in multiple variants of training code.

    In this paper, we present methodologies for efficient
    and effective logging practices for model training, with a focus on 
    techniques for hindsight logging. Our goal
    is for experienced model developers to learn and adopt these
    practices. To make this easier, we provide an open-source suite of tools for Fast
    Low-Overhead Recovery (\flor) that embodies our design across three 
    tasks: (i) efficient background logging in Python, (ii) adaptable 
    periodic checkpointing, and (iii) an instrumentation library that 
    codifies hindsight logging for efficient and automatic record-replay of 
    model-training. Model developers can use each \flor tool separately as
    they see fit, or they can use \flor in hands-free mode, entrusting it to 
    instrument their code end-to-end for efficient record-replay. 
    Our solutions leverage techniques from
    physiological transaction logs and recovery in database systems. 
    Evaluations on modern ML benchmarks demonstrate that \flor can 
    produce fast checkpointing with small user-specifiable overheads (e.g. 7\%), 
    and still provide hindsight log replay times orders of magnitude 
    faster than restarting training from scratch.

\end{abstract}

\maketitle

\pagestyle{\vldbpagestyle}
\begingroup\small\noindent\raggedright\textbf{PVLDB Reference Format:}\\
\vldbauthors. \vldbtitle. PVLDB, \vldbvolume(\vldbissue): \vldbpages, \vldbyear.\\
\href{https://doi.org/\vldbdoi}{doi:\vldbdoi}
\endgroup
\begingroup
\renewcommand\thefootnote{}\footnote{\noindent
This work is licensed under the Creative Commons BY-NC-ND 4.0 International License. Visit \url{https://creativecommons.org/licenses/by-nc-nd/4.0/} to view a copy of this license. For any use beyond those covered by this license, obtain permission by emailing \href{mailto:info@vldb.org}{info@vldb.org}. Copyright is held by the owner/author(s). Publication rights licensed to the VLDB Endowment. \\
\raggedright Proceedings of the VLDB Endowment, Vol. \vldbvolume, No. \vldbissue\ %
ISSN 2150-8097. \\
\href{https://doi.org/\vldbdoi}{doi:\vldbdoi} \\
}\addtocounter{footnote}{-1}\endgroup

\ifdefempty{\vldbavailabilityurl}{}{
\vspace{.3cm}
\begingroup\small\noindent\raggedright\textbf{PVLDB Artifact Availability:}\\
The source code, data, and/or other artifacts have been made available at \url{\vldbavailabilityurl}.
\endgroup
}

\section{Introduction}\label{sec:intro}

Due to the growing scale and complexity of sophisticated models~\cite{dean2012large,ratner2019sysml, bitter},
exploratory model development increasingly poses data management problems~\cite{stoica2017berkeley}.
At every step of exploration, model developers routinely track and visualize time series data to diagnose common training problems 
such as exploding/vanishing gradients~\cite{hochreiter1998vanishing}, dead ReLUs~\cite{lu2019dying}, and reward hacking~\cite{amodei2016concrete}.
Model developers use state-of-the-art loggers specialized to machine learning (e.g. TensorBoard~\cite{tensorboard}, and WandB~\cite{wandb})
to efficiently trace and visualize data as it changes over time.
The following are common examples of times series data logged in model training:
\begin{itemize}
    \item \textbf{Training Metrics}: The loss, accuracy, learning rate, and other metrics as they change over time.
    \item \textbf{Tensor Histograms}: Histograms of weights, gradients, activations, and other tensors as they change over time.
    \item \textbf{Images \& Overlays}: Segmentation masks, bounding boxes, embeddings, and other transformed images as they change over time.
\end{itemize}
In our experience, all model developers log some training metrics by default.
Whether their logging practice is characteristically conservative or optimistic 
depends on whether they log additional training data by default.
Next, we will illustrate the relevant differences between conservative and optimistic logging,
by introducing three archetypal characters: Mike for methodical conservative logging, Chuck for ad-hoc optimistic logging, and Judy for methodical optimistic logging.

\subsection{Conservative Logging}\label{subsec:intro_lo_overhead}

Conservative logging is eager and characterized by stable 
expectations about what data (and how much of it) will be necessary
for analysis~\cite{yuan2012conservative}. 
It is especially well-suited to later stages of the machine learning lifecycle,
where models are periodically trained for many hours on fresh data~\cite{strubell2019energy},
and refinements of the model training pipeline are usually light and limited to hyper-parameter tuning~\cite{garcia2018context}.

\subsubsection{\textbf{Mike} records everything (mnemonic for microphone)}\label{subsec:mike}
\begin{figure}
    \includegraphics[width=\columnwidth]{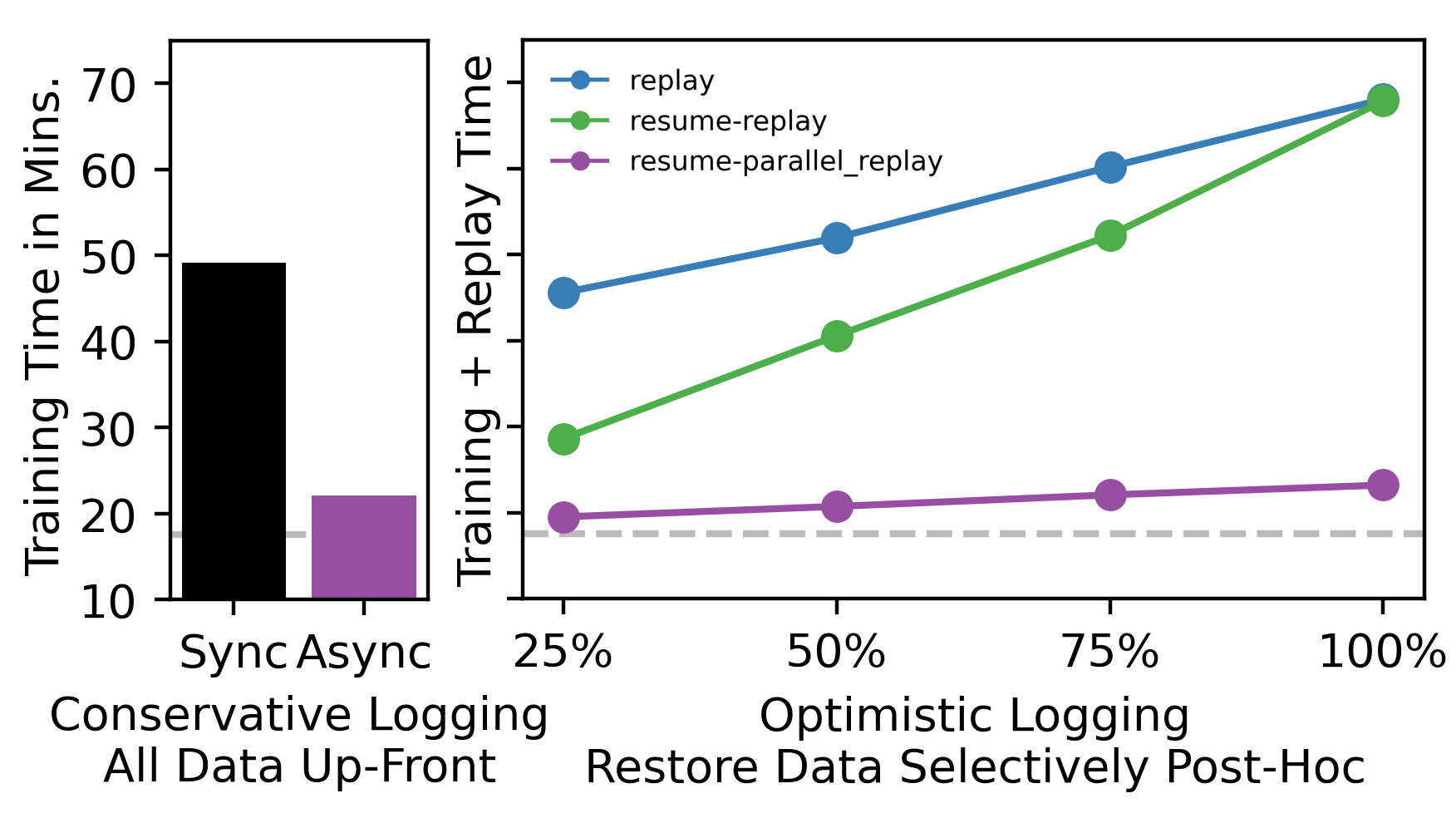}
    \caption{Conservative v. Optimistic logging performance at 100 epochs of Squeezenet on CIFAR-100~\cite{iandola2016squeezenet}.
    All workloads log tensor histograms for the activations, weights, and gradients 4$\times$ per epoch.
    Gray horizontal line corresponds to the same training job but without any logging. 
    Both, the purple histogram (conservative) and purple line (optimistic) correspond to \texttt{flor} logging.}
    \label{fig:mike}
\end{figure}

Mike is a model developer for a major tech company.
His organization's policy is that model developers should log training metrics, 
tensor histograms, and some images and overlays by default.
Although his logging practices can add substantial overhead to training (black bar in Figure~\ref{fig:mike}),
his jobs usually run as offline batches, and his productivity is not blocked on the results of training.
Moreover, when he receives an alert from the training or monitoring system,
the execution data he needs for post-hoc analysis will be ready.

Although reducing logging overhead is not a high priority for Mike, he is not the only developer
in his organization using high-end GPU clusters. 
At scale, even minor improvements to training efficiency will translate into measurable benefits
for the organization.
Later in this paper, we will present a tool for \textbf{low-overhead materialization in the background (Section~\ref{subsec:fork_logger})}.
Our tool enables Mike to continue to log data at the same rate and with his logger of choice (e.g. \texttt{tensorboardx}),
at a fraction of the original overhead (purple bar in Figure~\ref{fig:mike}).

\subsection{Optimistic Logging}\label{subsec:intro_lo_latency}

In contrast to conservative logging,
optimistic logging is an agile and lazy practice especially well-suited to 
early and unstructured stages of exploratory model development.
In optimistic logging, 
model developers log training metrics such as the loss and accuracy by default,
and defer collection of additional data until analysis time, when they may restore it selectively.
Execution data is restored by adding logging statements to training post-hoc, and replaying---possibly from checkpoint.
We refer to this practice as \emph{hindsight logging}.
Optimistic logging consists of (i) logging some training metrics by default, 
and (ii) selectively restoring additional training data post-hoc with hindsight logging.

Model developers gain agility in exploration from optimistic logging in three ways:
\begin{itemize}
    \item \textbf{Deferred Overhead}: Each training batch executes and produces results as quickly as possible. 
    Exploration time is finite, and the more alternatives a model developer can explore, the better.
    \item \textbf{Flexible Cost Schedule}: Model developers can selectively restore \emph{just} the data they need post-hoc.
    The fewer epochs they need to probe; the fewer resources they burn.
    \item \textbf{Separation of Concerns}: Concerns about what data to log and how much of it
    do not burden the developer during design and tuning. These considerations are postponed until analysis time.
\end{itemize}
In the fast path of the common case,
model developers get all the relevant information from the training loss, and move on.
In exceptional cases, however, training replay may be necessary for post-hoc data restoration.

\subsubsection{\textbf{Chuck} doesn't record anything (mnemonic for toss)}
Chuck is a first year graduate student in Astronomy who
is knowledgeable about the latest developments in machine learning,
but ill-versed in software engineering practices.
Chuck logs the training loss and accuracy by default,
but does not save checkpoints during training.

\subsubsection{\textbf{Judy} uses good judgment (mnemonic for judge)}
Judy is an experienced model developer with a strong software engineering background.
Like Chuck, Judy only logs the training loss and accuracy by default;
additionally, she checkpoints training periodically, 
and manages the numerous versions of her code, data, and checkpoints.

When either Chuck or Judy need to restore training data post-hoc---say,
the tensor histograms for the gradients at a rate of 4$\times$ per epoch---they will selectively
add the necessary logging statements to their training code, and re-train.
In many cases, Chuck and Judy will only want to restore data for a small fraction of training (e.g. 25\% of the epochs),
near the region where the loss exhibits an anomaly (e.g. near the middle of training).
Because Judy checkpointed her training execution, she is able to resume training at an arbitrary epoch.
Chuck, on the other hand, must retrain from the start.
In the right pane of Figure~\ref{fig:mike}, we plot training plus replay times for Chuck (blue line) and Judy (green line).

In this paper, we will concretely define the methodology that enables Judy to 
achieve effective record-replay of model training (Section~\ref{sec:methodical}).
Our goal is for experienced model developers to learn and adopt these best practices, for their numerous benefits.
One surprising consequence of Judy's approach is that she can \textbf{parallelize 
replay of model training} with her periodic checkpoints (Section~\ref{subsec:checkpointresume}).
The purple line in Figure~\ref{fig:mike} represents Judy's parallel replay.

Additionally, we evaluate (Section~\ref{sec:evaluation}) and open-source~\cite{flor_software} our Fast Low-Overhead Recovery suite (abbreviated as \flor)
for hindsight logging---with the following set of tools:
\begin{itemize}
    \item An optimized materialization library for low-overhead logging and checkpointing in Python (Section~\ref{subsec:fork_logger}).
    \item An adaptable periodic checkpointing mechanism to control \record{} overhead, so it never exceeds a user-specifiable limit (Section~\ref{subsec:adaptive_periodic}).
    \item An instrumentation library that can transparently transform Python training code to conform with the methodical hindsight logging approach (Section~\ref{subsec:instrumentation}).
    This protects model developers from incurring technical debt as a consequence of lapses in discipline, 
    and enables novices to restore time series data as efficiently as experts.
\end{itemize}
Model developers may use each \flor tool separately as they see fit, 
or they may use \flor in hands-free mode, entrusting it to instrument their code end-to-end for efficient record-replay.


\section{Methodical Hindsight Logging}\label{sec:methodical}

In hindsight logging, model developers can choose what to log long after model training: at analysis time and with a question in mind.
In essence, we want to query past execution state, without versioning that state in full. 
We draw inspiration from the rich body of work in databases dedicated to fast recovery~\cite{mohan1992aries, weikum2001transactional, zheng2014fast}.
Although that work focuses mostly on transactions, the lessons and trade-offs transfer naturally to execution recovery for arbitrary programs.
There are two means for recovering execution data: physically, by reading it from disk; and logically, by recomputing it. 
Both a purely physical approach and a purely logical approach are unattractive in our setting,
due to prohibitive overhead on record and prohibitive latency on replay, respectively.
Instead, hindsight logging---like transaction logging---embraces 
a hybrid ``physiological''~\cite{gray1992transaction} approach that takes partial checkpoints 
on the first pass (henceforth the \textit{record} phase), and uses those checkpoints 
to speedup redo (henceforth the \textit{replay} phase).

In this section, we give a high-level overview of the enabling methodology behind efficient hindsight logging:
\begin{enumerate}
    \item First and foremost, \textbf{checkpoint periodically} during training. 
    At least once per epoch for partial replay, but much less frequently is sufficient for parallel replay.
    \item Additionally, \textbf{enclose long-running code inside a conditional statement to exploit memoization} speedups.
    On record, the execution materializes the side-effects of each memoized block.
    On replay, model developers will run their code from the beginning without modification,
    and the execution will intelligently skip the recomputation of some blocks by loading their side-effects from memoization storage.
    \item Finally, include logic to \textbf{resume training from a checkpoint}.
    Replay of model training is embarrassingly parallel given periodic checkpoints.
    To parallelize replay of model training, a model developer dispatches multiple training jobs in parallel,
    each loading checkpoints to resume training from a different epoch and terminating early.
\end{enumerate}
If at any point through our forthcoming discussion the programming burden seems too high,
the reader should note that we also provide a tool that codifies and automatically applies these methods 
for the benefit of the user: an instrumentation library
that features a hands-free mode for convenience and robustness (Section~\ref{subsec:instrumentation}).

\subsection{Periodic Checkpointing \& Memoization}\label{subsec:skipblock}

\begin{figure}[t]
    \resizebox{0.97\columnwidth}{!}{
\begin{lstlisting}[style=customc]
init_globals()
checkpoint_resume(args, (net, optimizer))
for epoch in range(args.start, args.stop):
  if skipblock.step_into(...):
    for batch in training_data:
      predictions = net(batch.X)
      avg_loss = loss(predictions, batch.y)
      avg_loss.backward()
      optimizer.step()
  skipblock.end(net, optimizer)
  evaluate(net, test_data)
  
\end{lstlisting}}
    \caption{Training script prepared for methodical hindsight logging: checkpoint resume (line 2),
    block memoization (lines 4 - 10),
    and periodic checkpointing (line 10).
    The semantics of \skipblock are covered in subsection~\ref{subsec:skipblock}.}
    \label{fig:skipblock}
\end{figure}

Many model developers already checkpoint training periodically.
This is traditionally done for warm-starting training as well as for fault tolerance.
In this section, we show how to exploit further benefits from
periodic checkpointing, without incurring additional overheads.

In Figure~\ref{fig:skipblock}, we provide an example of how a model developer
would materialize the model and optimizer state once per epoch (line 10). 
This state serves a dual purpose. 
First, it comprises the relevant side-effects of the preceding code block (lines 4-9),
so it serves a
memoization purpose (computation skipping).
Second, it captures all of the state that is modified every epoch of training,
so it comprises a valid and complete checkpoint.
This dual purpose of selective state capture 
is a fortunate coincidence that arises naturally from the nested loops structure of model training.

We make use of the \skipblock language construct~\cite{chasins2017skip},
to denote block memoization.
The first two requirements for efficient record-replay are 
periodic checkpointing and block memoization.
Both are achievable by the following functionality,
which is encapsulated by the \skipblock for ease of exposition:
\begin{itemize}
    \item \textbf{Parameterized Branching}: \skipblock always applies the side-effects
    of the enclosed block to the program state, but does so in one of two ways: (a) by executing the enclosed block,
    or (b) by skipping the block
    and instead loading the memoized side-effects from its corresponding checkpoint.
    \skipblock automatically determines whether to execute or skip the enclosed block.
    It is parameterized by relevant execution state:
    i.e. \record{} execution, \replay{} resume, \replay{} execution, and whether the enclosed block is probed.
    \item \textbf{Side-Effect Memoization (i.e. Periodic Checkpointing)}: When the enclosed block is executed,
    \skipblock materializes its side-effects (the arguments passed to the call in line 10, Figure~\ref{fig:skipblock}).
    It is possible to optimize the \skipblock for low-overhead background materialization (Section~\ref{subsec:fork_logger}),
    and adaptable periodic materialization (Section~\ref{subsec:adaptive_periodic}), but these optimizations do not alter the semantics of \skipblock. 
    \item \textbf{Side-Effect Restoration}: Whenever the enclosed block is skipped,
    \skipblock restores its side-effects from its corresponding checkpoint (line 10, Figure~\ref{fig:skipblock}).
    \skipblock is able to efficiently locate an execution's corresponding checkpoint on disk,
    and apply its side-effects to the program state.
\end{itemize}

A block may not be skipped on \replay if the model developer adds a hindsight logging statement inside the block.
Although \skipblock memoizes the block's final state (that is visible to subsequent statements),
it does not materialize intermediate state, such as the activations of the model (line 6 in Figure~\ref{fig:skipblock}).
Consequently, if the model developer wishes to restore the model activations post-hoc,
it will not be possible to skip the nested training loop.
In such cases, parallel replay is the only option for reducing the latency of hindsight logging.

\subsection{Parallel Replay by Checkpoint Resume}\label{subsec:checkpointresume}

As we saw in the previous subsection,
our approach cannot avoid expensive recomputation
when intermediate training state, such as the gradients or activations
are logged post-hoc.
In such cases, model developers will want to reduce replay latency
by utilizing additional resources---specifically, more GPUs for parallelism.
Although auto-parallelizing arbitrary sequential code remains an open challenge~\cite{apostolakis2020perspective},
the replay of checkpointed model training is a special case:
training replay is embarrassingly parallel given periodic checkpoints.

As we multi-purposed periodic checkpointing in the previous section for memoization, so too
we now multi-purpose checkpoint resume---a current staple in the training code of many model developers---for parallel replay.
Parallel replay enables us to substantially cut hindsight logging latency, and due to the prevalence of checkpoint resume,
this is possible without incurring a programming burden.
To parallelize replay, a model developer simultaneously resumes training from various checkpoints:
\begin{enumerate}
    \item First, a model developer dispatches multiple training jobs (in parallel)
    \item Then, each job loads the checkpoint (line 2 in Figure~\ref{fig:skipblock}) that corresponds to the epoch it is resuming from.
    For example, to resume training at epoch 25, the job loads the checkpoint stored at the end of epoch 24.
    \item Finally, each job independently works on its share of work (see the \texttt{range} in line 3 of Figure~\ref{fig:skipblock}).
\end{enumerate}

\begin{figure}[t]
    \resizebox{0.9\columnwidth}{!}{
\begin{lstlisting}[style=customc]
init_globals()
for epoch in range(0, args.stop):
  if skipblock.step_into(... 
      && epoch >= args.start):
    for batch in training_data:
      predictions = net(batch.X)
      avg_loss = loss(predictions, batch.y)
      avg_loss.backward()
      optimizer.step()
  skipblock.end(net, optimizer)
  lr_scheduler.step()
  evaluate(net, test_data)
  
\end{lstlisting}}
    \caption{Training script prepared for methodical hindsight logging. Training can resume from a partial checkpoint (no \texttt{lr\_scheduler} in checkpoint).}
    \label{fig:pseudoresume}
\end{figure}

\subsubsection{\textbf{Pseudoresuming} from partial checkpoints}\label{subsec:pseudoresume}
When model developers write code for periodic checkpointing themselves,
they can ensure that the objects they capture constitute a complete checkpoint.
However, when model developers entrust \flor to instrument their code for automatic periodic checkpointing,
\flor will not be able to automatically determine whether the checkpoint is complete or partial with respect to training. 
As we will discuss in Section~\ref{subsec:instrumentation}, \flor can only estimate the side-effects 
of blocks of code enclosed by \skipblock{}s:
a restriction we use to render our static analysis tractable.
\flor will not estimate the side-effects of the program at arbitrary points,
and it will not check whether the data materialized constitutes a complete (or partial) checkpoint with respect to training,
since doing so statically (i.e. with low overhead) would be intractable in Python~\cite{Holkner2009EvaluatingTD, spuler1994compiler,nicolay2015detecting}.

Consequently, \flor assumes checkpoints materialized automatically are partial with respect to training.
By partial, 
we mean that there are objects modified every epoch 
that are not stored by the checkpoints (e.g. the \texttt{lr\_scheduler} in Figure~\ref{fig:pseudoresume}).
As a result, it is not possible to resume training 
from an arbitrary epoch merely by loading a partial checkpoint (i.e. a physical recovery approach).
Instead, we start training from the beginning (line 2 in Figure~\ref{fig:pseudoresume}), 
and use the partial checkpoints
to skip recomputation of memoized blocks during the initialization --- or 
\emph{pseudoresume} --- phase (lines 3-4 in Figure~\ref{fig:pseudoresume}). 
This approach is characteristically physiological because it 
relies on a combination of recomputation and disk reads for recovery.
Although \emph{pseudoresume} is especially important for auto-parallelizing replay of model training,
we share this method here because novice model developer may accidentally store partial checkpoints.
This technique allows them to resume training efficiently all the same.

For illustration, suppose, that the model developer wants to resume training from epoch 25,
using the script in Figure~\ref{fig:pseudoresume}.

\noindent\textbf{Pseudoresume phase} for epochs in the range 0-24 (inclusive):
\begin{enumerate}
    \item \skipblock skips the nested training loop (lines 3-4 in Figure~\ref{fig:pseudoresume}).
    \item \skipblock loads the side-effects of the nested training loop from disk (line 10 in Figure~\ref{fig:pseudoresume}).
    \item All other statements execute normally.
\end{enumerate}
\noindent\textbf{Execution phase} from epoch 25 onward:
\begin{enumerate}
    \item \skipblock enters (steps into) the nested training loop (lines 3-4 in Figure~\ref{fig:pseudoresume}).
    \item All other statements execute normally.
\end{enumerate}

In summary, memoization can be used to resume model training from an arbitrary epoch, even in the absence of complete checkpoints.
As we will show in the evaluation, the overhead of \textit{pseudoresume} is amortized in parallel replay,
so that the difference between checkpoint resume and \textit{pseudoresume} is imperceivable to the end-user.
This result is important because it enables us to efficiently auto-parallelize the replay of model training, even without complete checkpoints.


\section{Tooling for Hindsight Logging}\label{sec:tooling}

We provide a suite of \textit{Fast Low-Overhead Recovery} tools---\flor for short---as aid to the developer.
Model developers may use each tool separately as they see fit, 
or they may use \flor in hands-free mode, entrusting it to instrument their code end-to-end for efficient record-replay.
\flor provides the following tools:
\begin{itemize}
    \item An optimized materialization library for low-overhead logging and checkpointing (Section~\ref{subsec:fork_logger}).
    \item An adaptable periodic checkpointing mechanism to control \record{} overhead, so it never exceeds a user-specifiable limit (Section~\ref{subsec:adaptive_periodic}).
    \item An instrumentation library that can transparently transform training code to conform to the methodical hindsight logging approach (Section~\ref{subsec:instrumentation}).
    This protects model developers from incurring technical debt as a consequence of lapses in discipline, 
    and enables novices to restore time series data as efficiently as experts.
\end{itemize}

\subsection{Background Logging}\label{subsec:fork_logger}

\flor provides a background materialization mechanism optimized for PyTorch,
which is compatible with model developer's machine learning logging service of choice 
(for example, TensorBoard~\cite{tensorboard}, MLFlow~\cite{zaharia2018accelerating}, and WandB~\cite{wandb}). 
Background logging is used natively by \skipblock{}s for low-overhead periodic checkpointing (Section~\ref{subsec:skipblock}).
It is also available separately as a library for end-users.

Both logging and checkpointing can add measurable overhead to training 
because they require moving data from GPU memory to DRAM,
serializing it into byte arrays, 
and then writing those arrays to disk. 
Of the latter two, serialization is typically much more expensive than I/O: 
by an average factor of 4.3$\times$ according to our microbenchmarks~\cite{Liu:EECS-2020-79}. 
Consequently, after copying select data from GPU memory to DRAM (so it is protected from overwrites),
we would like to take materialization (both serialization and I/O) off the main thread---which is dedicated to model training---and 
do it in the background. 

Despite its maturity and widespread popularity, Python makes this very difficult.
The Python interpreter has a notorious Global Interpreter Lock 
that prevents parallelism among Python threads. 
Unfortunately,
the Python IPC schemes (e.g., the \texttt{multiprocessing} library) also require serialization by the sending process---returning us to our original problem. 
To avoid serialization we could use a solution like Apache Plasma, but
it only avoids serialization for a subset of Python data types (notably dataframes and arrays) and actually cannot serialize other data types including Pytorch tensors.
We eventually found a workaround at the operating system level, using
\texttt{fork()} as a mechanism to achieve efficient one-shot, one-way IPC between a parent and child process,
with copy-on-write concurrency.
To materialize a \record{} checkpoint, the main process forks and then immediately resumes model training; 
the child process serializes the checkpoint, writes it to disk, and then terminates.
To prevent too many calls to \texttt{fork()}, we buffer up checkpoints and process them in batches of 5000 objects.
Given the short lifespan of these child processes and an infrequent rate
of forking due to batching, we have never seen more than
two live children at any point in our evaluations---including in models that ran for many hours (Section~\ref{sec:evaluation}).

In a technical report~\cite{Liu:EECS-2020-79}, we provide a more detailed discussion 
of the design and performance of our background materialization mechanism. 
This mechanism cuts logging overheads by 73.5\% on average, according to our microbenchmarks~\cite{Liu:EECS-2020-79}.
Execution speedups due to background logging are modest
for workloads whose logging overheads are dominated by periodic checkpointing 
($\mu=4.76\%$ overhead down to $\mu=1.74\%$).
This is because periodic checkpointing is already light 
and doesn't add much overhead to training.
However, as we saw in Figure~\ref{fig:mike}, background logging
can have a drastic effect when used for conservative logging (180\% overhead down to 26\%),
since logging overheads account for a much larger fraction of end-to-end training times in those cases.

\subsection{Adaptable Periodic Checkpointing}\label{subsec:adaptive_periodic}

In this section, we present a decision rule for dynamically calculating an appropriate checkpointing \textit{period} or frequency.
This condition is automatically tested by \skipblock{}s to adapt the frequency of checkpointing to each training workload.
For many developers, checkpointing once per epoch is a good default,
but in general, the right checkpointing frequency depends on the training workload:
e.g. how fast or slow the code executes relative to the size of its checkpoints.
The goal of adaptable periodic checkpointing is to automatically materialize
checkpoints as frequently
as will increase expected replay speedups, subject to the constraint that record overhead
does not exceed a user-specifiable limit.

Next we derive the invariants we use for adaptable periodic checkpointing. 
We refer the reader to the notation in Table~\ref{tab:symtabl}.

\begin{table}
    \centering
    \caption{Symbol table for Adaptable Periodic Checkpointing}
    \resizebox{\columnwidth}{!}{%
    \begin{tabular}{|l|l|}
    \hline
    Symbol        & Description                  \\ \hline \hline
    $M_i$         & time to materialize side-effects of  block identified by $i$   \\ \hline
    $R_i$         & time to restore side-effects of block identified by  $i$   \\ \hline
    $C_i$         & time to compute block identified by $i$                       \\ \hline
    $n_i$         & count of executions (so far) for block $i$         \\ \hline
    $k_i$         & count of checkpoints (so far) for block $i$    \\ \hline
    $G$           & degree of \replay{} parallelism                                       \\ \hline
    $c$           & constant scaling factor                                       \\ \hline
    $\epsilon$    & tunable parameter denoting overhead tolerance                  \\ \hline
    \end{tabular}
    }
    \label{tab:symtabl}
    \end{table}

\subsubsection{\textbf{The Record Overhead Invariant}} 
We require that the materialization overhead of a block is 
at most a small fraction of its computation time: $M_{i} < \epsilon C_{i}$. 
This simplistic invariant is enough to ensure that \record{} 
never exceeds a user-specifiable overhead ($\epsilon$),
but it is all-or-nothing: a block is memoized always or never.
Since blocks are often nested inside loops,
and model developers may parallelize \replay{} even with a small number of checkpoints (e.g. 2 checkpoints: 3$\times$ parallelism),
we need to relax our invariant to account for periodic checkpointing.
Specifically, due to the nested loops structure of model training, we introduce $n_i$ and $k_i$, as follows: 
\begin{equation} \label{eq:lo_ovrhd}
\begin{split}    
k_i M_{i} &< n_{i} \epsilon  C_{i}  \quad
 \Rightarrow \quad \frac{M_i}{C_i} < \frac{n_i \epsilon}{k_i}
\end{split}
\end{equation}

\subsubsection{\textbf{The Replay Latency Invariant}} 
To avoid regret, \record{}-\replay{}
should always be faster than two vanilla executions (with neither overhead nor speedups).
Even for hindsight logging workloads that do not permit partial replay,
the speedups from parallel replay alone should more-than-offset the overhead incurred on \record{}.
Accounting for \record{} overhead, we can assess each block $i$ for this condition as follows:  
\begin{equation}\label{eq:leastbenefit}
    M_{i} + R_{i} + \left(\frac{n_{i}}{G} - 1 \right) C_{i} < n_{i} C_{i}
\end{equation}
The $-1$ in Equation~\ref{eq:leastbenefit} accounts for the fact that each parallel worker resumes from a stored checkpoint and does not need to compute its first iteration.

Because $G$ is determined on \replay{} and is not known during \record{}, 
we satisfy the Replay Latency Invariant by testing Equation~\ref{eq:lolat} instead.
Equation~\ref{eq:lolat} guarantees the Replay Latency Invariant as long as there is some parallelism ($G \geq 2$); we omit the details for brevity. 
\begin{equation} \label{eq:lolat}
\begin{split}
M_{i} + R_{i} < \frac{n_{i}}{k_{i}} C_{i} \quad \text{and} \quad
R_{i} = c M_{i}  \\
\quad \Rightarrow  \frac{M_{i}}{C_i} <  \frac{n_{i}}{k_{i} (1+c)} 
\end{split}
\end{equation}
Because the time to restore is not known at \record{} time, we assume that it is proportional to the time to materialize. 
Our naive assumption is $c=1.0$, and this estimate is refined after observing
materialization and restoration times from \record{}-\replay{}. 
In our case, the average scaling factor over all measured workloads (Table~\ref{tab:experiment_list}) turned out to be $c=1.38$. 

\subsubsection{\textbf{The Joint Invariant}}\label{subsec:ratio_test}
The Joint Invariant is automatically checked by \skipblock{}s 
at \record{} time for adapting the frequency of checkpointing.
Blocks are tested after executing, but before materialization.
By restricting memoization to blocks that pass the Joint Invariant test,
\texttt{flor} simultaneously satisfies the Record Overhead and Replay Latency invariants.
This follows from the fact that the Joint Invariant is derived by
algebraic manipulation of the two invariants.
\begin{equation} \label{eq:adaptchkpt}
    \begin{split}
    \frac{M_{i}}{C_{i}} &< \frac{n_{i}}{k_{i} + 1}  \min{(\frac{1}{1+c}, \epsilon)} \\
    c &= 1.38, \epsilon=0.0667
    \end{split}
    \end{equation}
Note the $k_i + 1$ in Equation~\ref{eq:adaptchkpt}: 
this accounts for the fact that the test is performed after 
the execution of the block but before the materialization of its checkpoint.
The goal is for the invariant to continue to hold if the checkpoint is materialized.
We derive the Joint Invariant, Equation~\ref{eq:adaptchkpt}, from 
Equation~\ref{eq:lo_ovrhd} and Equation~\ref{eq:lolat}.
Both invariants are satisfied when the computed ratio, $M_{i} / C_{i}$, 
is less than the minimum of both thresholds. 

\subsection{Instrumentation for Hands-Free Mode}\label{subsec:instrumentation}

As desired by the user, \flor can instrument their model training code 
for automatic and efficient record-replay. 
The principal objectives of \flor instrumentation are twofold:
\begin{enumerate}
    \item Memoization and periodic checkpointing by nesting loops inside a \skipblock, and then statically estimating their side-effects.
    \item Auto-parallelization of training replay by a syntax-directed transformation of loop iterators, 
    enabling \textit{pseudoresume} from partial checkpoints (Subsection~\ref{subsec:pseudoresume}).
\end{enumerate}

\begin{table}[]
\centering
\caption{Set of rules for static side-effect analysis. 
At most one rule is activated by each program statement.
The rules are sorted in descending order of precedence.
}
\resizebox{\columnwidth}{!}{%
\begin{tabular}{|l|l|l|}
\hline
Rule & Pattern                                          & $\Delta$Changeset                  \\ \hline \hline
0 & $v_1, ..., v_n = u_1, ..., u_m \wedge \exists v_i \in \texttt{Changeset}$          & \texttt{No Estimate}  \\ \hline
1 & $v_1, ..., v_n = obj.method(arg_1, ..., arg_m)$     & $\{obj, v_1, ..., v_n\}$    \\ \hline
2 & $v_1, ..., v_n = func(arg_1, ..., arg_m)$           & $\{v_1, ..., v_n\}$  \\ \hline
3 & $v_1, ..., v_n = u_1, ..., u_m$           & $\{v_1, ..., v_n\}$  \\ \hline
4 & $obj.method(arg_1, ..., arg_m)$     & $\{obj\}$    \\ \hline
5 & $func(arg_1, ..., arg_m)$           & \texttt{No Estimate}  \\ \hline

\end{tabular}
}
\label{tab:cdsm_examp}
\end{table}

\subsubsection{\textbf{Autorecording Model Training}}\label{subsec:side_effect_analysis}
The first goal of instrumentation is to efficiently and correctly memoize loop executions for 
the model developer---without their intervention.
\flor memoizes loops because, in machine learning,
they correspond to long-running code,
and unlike arbitrary ``long-running code'', loops can be detected statically.
Ensuring correct and efficient memoization requires 
(i) capturing all of the loop's side-effects,
and (ii) avoiding the capture of too many redundancies.
Unfortunately, due to the language's dynamic features and extensive reliance on (compiled) C extensions, 
an exact and efficient side-effect analysis in Python is 
intractable~\cite{Holkner2009EvaluatingTD, spuler1994compiler,nicolay2015detecting}.
Past work overcomes Python's analysis limitations 
by restricting the expressiveness of the language~\cite{ancona2007rpython, lam2015numba}, 
making some assumptions (e.g. that the variables don't change types~\cite{aycock2000aggressive}),
or relying on user source annotations~\cite{vitousek2014design}.
In a similar vein, we achieve efficient side-effect analysis
by assuming that loop bodies in model training are predominantly written in PyTorch~\cite{paszke2019pytorch}.
To the extent that loops deviate from our assumption, our static analysis will be unsafe 
(i.e. may misdetect side-effects), so we will automatically perform deferred correctness checks after replay
and report any anomalies to the programmer.
We find that our assumption holds frequently enough to be useful for hindsight logging purposes.

Model developers do not typically build models or write training algorithms from scratch.
Instead, they rely on popular machine learning frameworks such as PyTorch.
Like many 3rd-party libraries, PyTorch has a well-defined 
interface by which it modifies the user's program in limited ways~\cite{pytorch_docs}.
The effects of PyTorch on the user's program are limited to (i) assignments and (ii) encapsulated state updates from method calls.
As a result, all the side-effects of PyTorch code can be detected statically, 
with two notable exceptions: when an optimizer modifies a model, 
and when a learning rate scheduler modifies an optimizer~\cite{pytorch_docs_opt}.

First, \flor estimates a set of changes (``changeset'') for each block using the six rules in Table~\ref{tab:cdsm_examp}. 
\flor walks the abstract syntax tree statement by statement, testing which rule is activated by each statement.
The changeset for a block accumulates the individual changes of its member statements. 
Rules have a precedence such that at most one is activated per statement. Statements that activate no rule are ignored.

Next, \flor performs a filtering step on the changeset to remove variables that are scoped to the body of the loop.
\flor removes from the changeset any 
variable that is defined in the body of the loop (henceforth ``loop-scoped variable''), 
under the assumption that this variable is local to the loop and is not read after the end of the loop.

Finally, we make use of our encoded library-specific knowledge to augment the changeset at runtime 
(this is the only step that is not done statically).
For PyTorch, it suffices to encode two facts~\cite{pytorch_docs_opt}: 
(i) the model may be updated via the optimizer; and 
(ii) the optimizer may be updated via the learning rate schedule.
\flor augments the changeset to include side-effects which were not detected by the rules, 
but which can be inferred from other elements in the changeset.

\begin{figure}[t]
    \resizebox{0.9\columnwidth}{!}{
\begin{lstlisting}[style=customc]
def flor.generator(*args):
  pseudoresume_sgmnt, work_sgmnt = partition(*args)
  skipblock.set_state('replay', 'skip')
  for element in pseudoresume_sgmnt:
    yield element
  skipblock.set_state('replay', 'step_into')
  for element in work_sgmnt:
    yield element

for epoch in flor.generator(range(N), PID, NPARTS):
  ...
\end{lstlisting}}
    \caption{\flor instrumentation nests the main loop's iterator inside a generator to parallelize replay (with \textit{pseudoresume}). 
    A generator defines an iterator, 
    and enables us to control global state between iterations of the main loop.}
    \label{fig:generator}
\end{figure}

\subsubsection{\textbf{Autoparallelizing Replay}}\label{subsec:autoparallel_replay}
The second goal of instrumentation is to autoparallelize 
replay of model training, assuming partial checkpoints exist due to \skipblock{}s.
Replay instrumentation consists of wrapping the main loop's iterator
inside a \texttt{flor.generator} (line 10 in Figure~\ref{fig:generator}),
to model the \textit{pseudoresume} behavior we covered earlier (in Subsection~\ref{subsec:pseudoresume} and Figure~\ref{fig:pseudoresume}).
We can identify the main loop at replay time, by having measured loop execution times during record.
We define the longest-running loop as the main loop.
Generators define an iterator by a series of \texttt{yield} statements,
and allow us to control global program state between iterations of the main loop;
namely, toggle the \skipblock{}s from a \textit{skip} state to a \textit{step-into} state between epochs (lines 3, 6 in Figure~\ref{fig:generator}).
We implement parallel replay by having every parallel worker (\texttt{NPARTS} in total)
execute the same instrumented code (as in Figure~\ref{fig:generator}), and \flor sets \texttt{PID} to 
a different value for each worker so they work on distinct segments of the main loop.

\subsubsection{\textbf{Deferred Checks for Correctness}}\label{subsec:deferred_checks}
As we have discussed, 
Python's dynamic features and extensive reliance on (compiled) C extensions, 
make an exact and efficient side-effect analysis intractable.
\flor{}'s approach to detecting side-effects is efficient but unsafe: it 
may misdetect side-effects and thus fail to checkpoint sufficient state for correct replay. 
To mitigate risk, we automatically check that common user-observable state between \record{} and \replay{} matches~\cite{altekar2009odr}.
The standard training metrics that get logged by default (e.g. the loss and accuracy) form a fairly unique
fingerprint of a model's training characteristics, so it's hard to perturb state or data that the model depends on
without this being reflected in one of the model's metrics.
Consequently, at the end of \replay{}, we run \texttt{diff}, and warn the user if 
the \replay{} logs differ from the record logs in any way other than the statements added for hindsight logging. 

\section{Evaluation}\label{sec:evaluation}

\begin{table*}
    \centering
    \caption{Computer vision and NLP benchmarks used in our evaluation.}
    \resizebox{\textwidth}{!}{%
    \begin{tabular}{|l|l|l|l|l|l|l|}
    \hline
    Name & Benchmark  & Task                           & Model            & Dataset        & Train/Tune  & Epochs \\ \hline \hline
    RTE  & GLUE       & Recognizing Textual Entailment & RoBERTa          & RTE            & Fine-Tune   & 200    \\ \hline
    CoLA & GLUE       & Language Acceptability         & RoBERTa          & CoLA           & Fine-Tune   & 80     \\ \hline
    Cifr & Classic CV & Image Classification           & Squeezenet       & Cifar100       & Train       & 200    \\ \hline
    RsNt & Classic CV & Image Classification           & ResNet-152       & Cifar100       & Train       & 200    \\ \hline
    Wiki & GLUE       & Language Modeling              & RoBERTa          & Wiki           & Train       & 12     \\ \hline
    Jasp & MLPerf     & Speech Recognition             & Jasper           & LibriSpeech    & Train       & 4      \\ \hline
    ImgN & Classic CV & Image Classification           & Squeezenet       & ImageNet       & Train       & 8      \\ \hline
    RnnT & MLPerf     & Language Translation           & RNN w/ Attention & WMT16          & Train       & 8      \\ \hline
    \end{tabular}
    }
    \label{tab:experiment_list}
\end{table*}

To assess \flor{}'s ability to meet the goals of Section~\ref{sec:intro} in practice,
we evaluated eight diverse machine learning workloads, 
taken from three separate benchmarks: 
classic computer vision, 
the General Language Understanding and Evaluation (GLUE)~\cite{wang2018glue}, 
and ML Perf~\cite{mattson2019mlperf} (Table~\ref{tab:experiment_list}).
These workloads vary in their tasks, 
model architectures, execution time scales, 
and software engineering patterns;
they are jointly representative of a large class of model training workloads.
Every experiment was run on P3.8xLarge EC2 instances
with 4 Tesla V100 GPUs, 
64 GB of GPU memory in aggregate, 
32 vCPUs, 
244 GB of RAM, 
and an EBS bandwidth (IO throughput) of 7Gbps.
The checkpoints generated by \flor{} \record{} were spooled from EBS to an S3 bucket by a background process.

\subsection{Flor Record Overhead is Low}\label{subsec:eval_record_overhead}

\begin{figure}
    \includegraphics[width=\columnwidth]{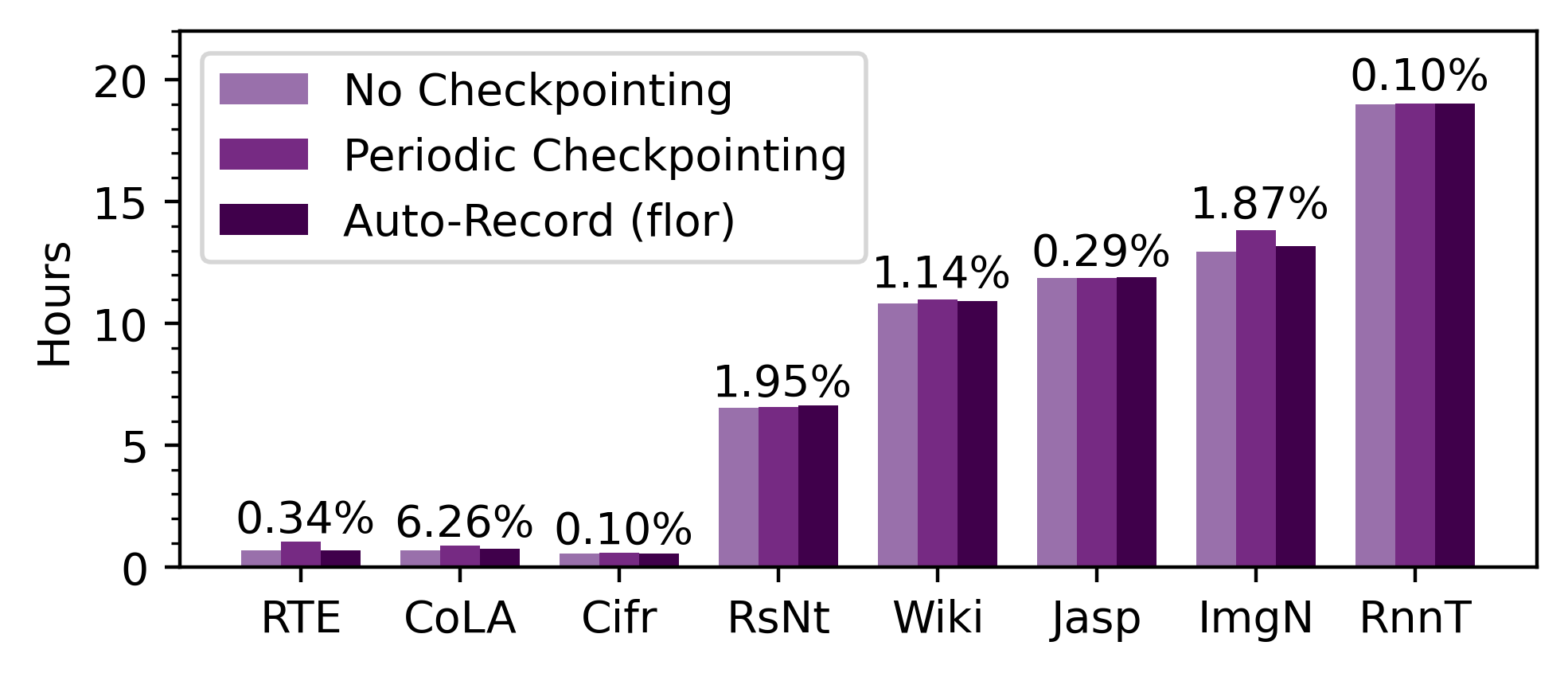}
    \caption{Comparison of model training times, with and without checkpointing, in hours. 
    ``Periodic Checkpointing'' measures the time achievable when a model developer judiciously selects the contents of a checkpoint, at a frequency of once per epoch.
    The overhead added by \flor Record is denoted by the text labels over each group of bars.}
    \label{fig:record_histograms}
\end{figure}

We compared the overhead added by manual periodic checkpointing, at a rate of once per epoch,
against the overhead added by automatic \flor record (Figure~\ref{fig:record_histograms}). 
We did not manually set the period for any of the \flor record experiments.
The checkpointing period was automatically calibrated by the mechanism in Section~\ref{subsec:adaptive_periodic}.
\flor record does not add significant overhead to training, so it may be enabled by default. 
Moreover, the \flor instrumentation library is able to achieve
the same effect as hand-tuned periodic checkpointing, with competitive performance, and without intervention from the user.
For manual periodic checkpointing, we assume that each checkpoint is complete with respect to training.
For \flor record we only assume partial checkpoints: each checkpoint corresponds to the side-effects
of the code block it memoizes (Section~\ref{subsec:skipblock}), but it may be incomplete with respect to training.

\botitem{Takeaway}: \flor record adds minor overhead, so it can be enabled by default.
A novice model developer who uses \flor can achieve similar performance to an experienced model developer.

\subsection{Flor Record Overhead is Adaptable}

\begin{figure}
    \centering
    \includegraphics[width=\columnwidth]{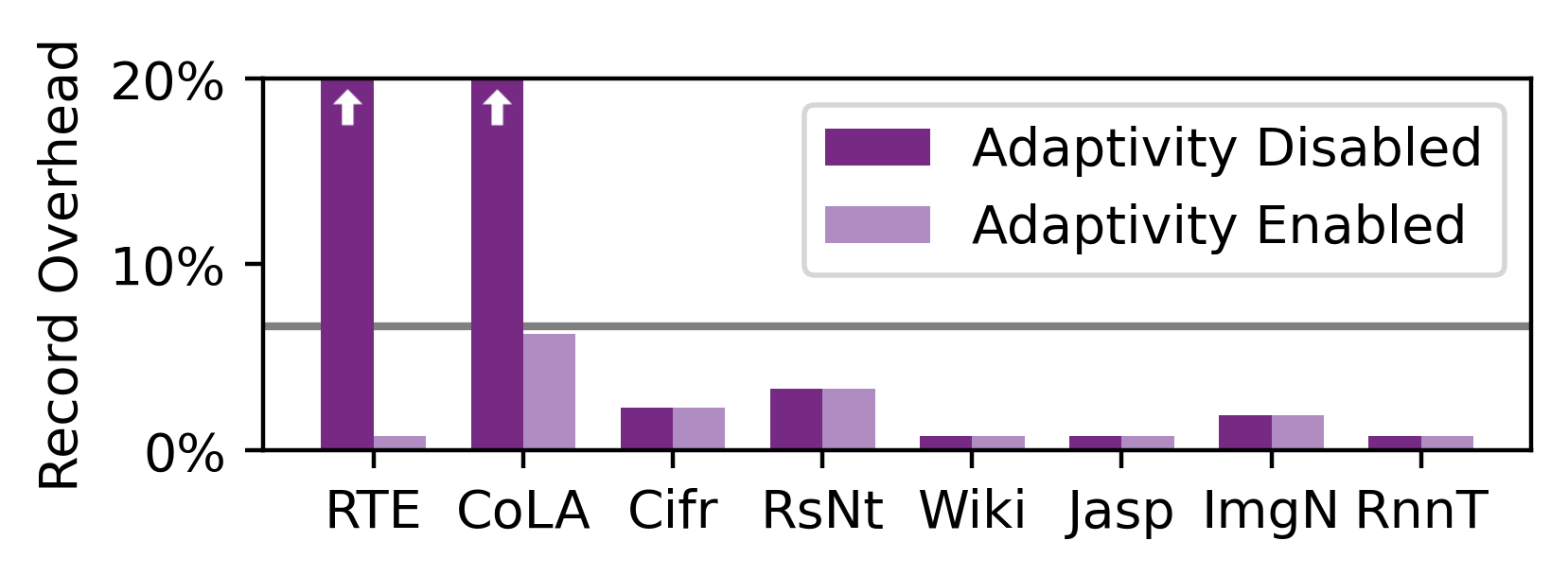}
    \caption{Impact of adaptivity on \flor record overhead.
    The two upward arrows denote extreme values: adaptivity-disabled overhead is 91\% for RTE and 28\% for CoLA. 
    The user-specifiable overhead tolerance (6.67\%) is denoted by the gray horizontal line.
    No workload exceeds the overhead limit with adaptive checkpointing.
    }
    \label{fig:adaptiveoverhead}
\end{figure}

Different model developers have different sensitivies to overhead.
In this section, we measured that \flor record is able to adjust its checkpointing frequency
to stay within the user-specifiable overhead limits (e.g. $\epsilon = 6.67\%$).
The nested training loops in most model training workloads are memoized 
every epoch by \flor{}'s adaptive checkpointing mechanism.
This is because the time to materialize their checkpoints is negligible compared to the time
it takes to execute them.
In contrast, the sharp drop in overhead for fine-tuning workloads
is due to their less frequent checkpointing (Figure~\ref{fig:adaptiveoverhead}) .
Fine-tuning workloads are checkpointed less frequently because their loops have poor materialization time to computation time ratios.
In other words, their checkpoints are massive relative to their short execution times.
This is the case because the vast majority of weights are frozen in model fine-tuning,
so a loop execution quickly updates a small fraction of values in an enormous model~\cite{howard2018universal}.

\botitem{Takeaway}: Adaptive checkpointing drastically reduces overhead on model fine-tuning workloads (RTE \& CoLA), 
and ensures that no workload exceeds the user's overhead tolerance.

\subsection{Flor Replay Latency is Low}\label{subsec:eval_latency}

\begin{figure*}
    \centering
    \begin{minipage}[t]{0.45\linewidth}
    \resizebox{\columnwidth}{!}{
\begin{lstlisting}[style=customc]
init_globals()
checkpoint_resume(args, (net, optimizer))
for epoch in range(args.start, args.stop):
  if skipblock.step_into(...):
    for batch in training_data:
      predictions = net(batch.X)
      avg_loss = loss(predictions, batch.y)
      avg_loss.backward()
      optimizer.step()
      tensorboard.add_histogram(net.params())
  skipblock.end(net, optimizer)
  evaluate(net, test_data)
  tensorboard.add_overlays(net, test_data)
\end{lstlisting}}
    \caption{Model training example with checkpoint resume. Lines 10 and 13 correspond to hindsight logging statements, 
    or logging statements added after training.}
    \label{fig:skipblocktb}
    \end{minipage}
    \quad \quad \quad
    \begin{minipage}[t]{0.45\linewidth}
    \resizebox{\columnwidth}{!}{
\begin{lstlisting}[style=customc]
init_globals()
for epoch in range(0, args.stop):
  if skipblock.step_into(... 
      && epoch >= args.start):
    for batch in training_data:
      predictions = net(batch.X)
      avg_loss = loss(predictions, batch.y)
      avg_loss.backward()
      optimizer.step()
      tensorboard.add_histogram(net.params())
  skipblock.end(net, optimizer)
  lr_scheduler.step()
  evaluate(net, test_data)
  tensorboard.add_overlays(net, test_data)
  
\end{lstlisting}}
    \caption{Model training example with checkpoint pseudoresume. Lines 10 and 14 correspond to hindsight logging statements, 
    or logging statements added after training.}
    \label{fig:skipblock_pseudotb}
    \end{minipage}
\end{figure*}

\begin{figure}[]
\centering
\includegraphics[width=\columnwidth]{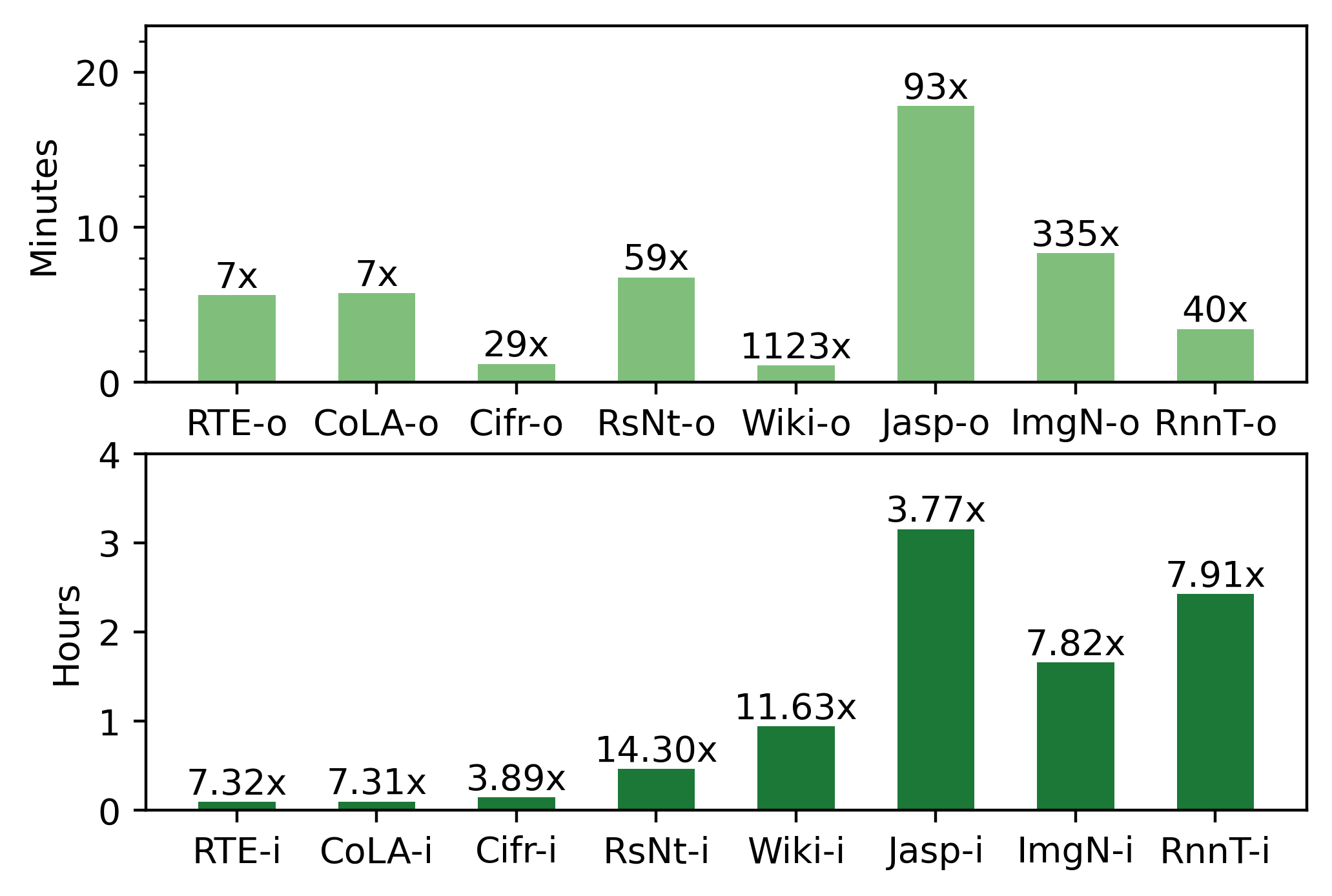}
\caption{Replay latency, factored by the position of hindsight logging statements.
The top plot reports partial and parallel replay speedups when
the model developer probes only the outer main loop (as in line 13 of Figure~\ref{fig:skipblocktb}).
The bottom plot reports parallel-only replay speedups when
the model developer probes the inner training loop and a full re-execution is needed (as in line 10 of Figure~\ref{fig:skipblocktb}).
Each workload uses as many machines, from a pool of four machines, as will result in parallelism gains.
Text labels show speedup factors relative to naive re-execution.
}
\label{fig:replay_histograms}
\end{figure}

In this section, we measure the replay speedups 
achieved by \flor replay, assuming \flor record checkpoints were materialized during training.
Consequently, we measure the replay speedups when \flor instruments 
model developers' code end-to-end for efficient hindsight logging---without intervention from the developer.

Replay latencies are query dependent: 
they depend on the position of hindsight logging statements in the code.
In cases when the model developer probes only the outer loop of training (as in line 13 of Figure~\ref{fig:skipblocktb}),
partial \replay{} can provide  latencies on the order of minutes, even when model training takes many hours to execute.
This is achieved by skipping unnecessary recomputation with loop memoization (e.g. skipping the nested training loop).
The top subplot in Figure~\ref{fig:replay_histograms} shows outer-loop probe latencies for each of our models. 
Note the improvements range from $7\times$ to $1123\times$---with the more significant 
improvements favoring the longer experiments (recall Figure~\ref{fig:record_histograms}).

When the model developer logs data post-hoc from the inner training loop (as in line 10 in Figure~\ref{fig:skipblocktb}),
then that loop must be re-executed on \replay{}, and it will not 
contribute to savings from loop memoization.
For these workloads, we will need to rely on parallelism to reduce latencies.
We measured the hindsight logging latencies when a full re-execution of
model training was necessary by running \replay{} on 
multiple machines---this is shown in the bottom subplot in Figure~\ref{fig:replay_histograms}). 

The parallel replay workloads used as many machines from the pool of 4  machines
as would provide further parallelism gains. Each machine has 4 GPUs. 
In the limit, every epoch may re-execute in parallel, but the degree of
parallelism may be increased even further by checkpointing additional state, which we leave as future work.

\textbf{Takeaway}: Assuming no work or guidance from the model developer, beyond the insertion of a couple
hindsight logging statements, \flor automatically parallelizes and conditionally skips computation on the re-execution of 
model training (Section~\ref{subsec:autoparallel_replay}).
It achieves the greatest replay speedups from partial replay, but may scale out replay to multiple machines for further speedups.

\subsection{Ideal Parallelism and Scale-out}\label{subsec:scaleout}

\begin{figure}
    \centering
    \includegraphics[width=\columnwidth]{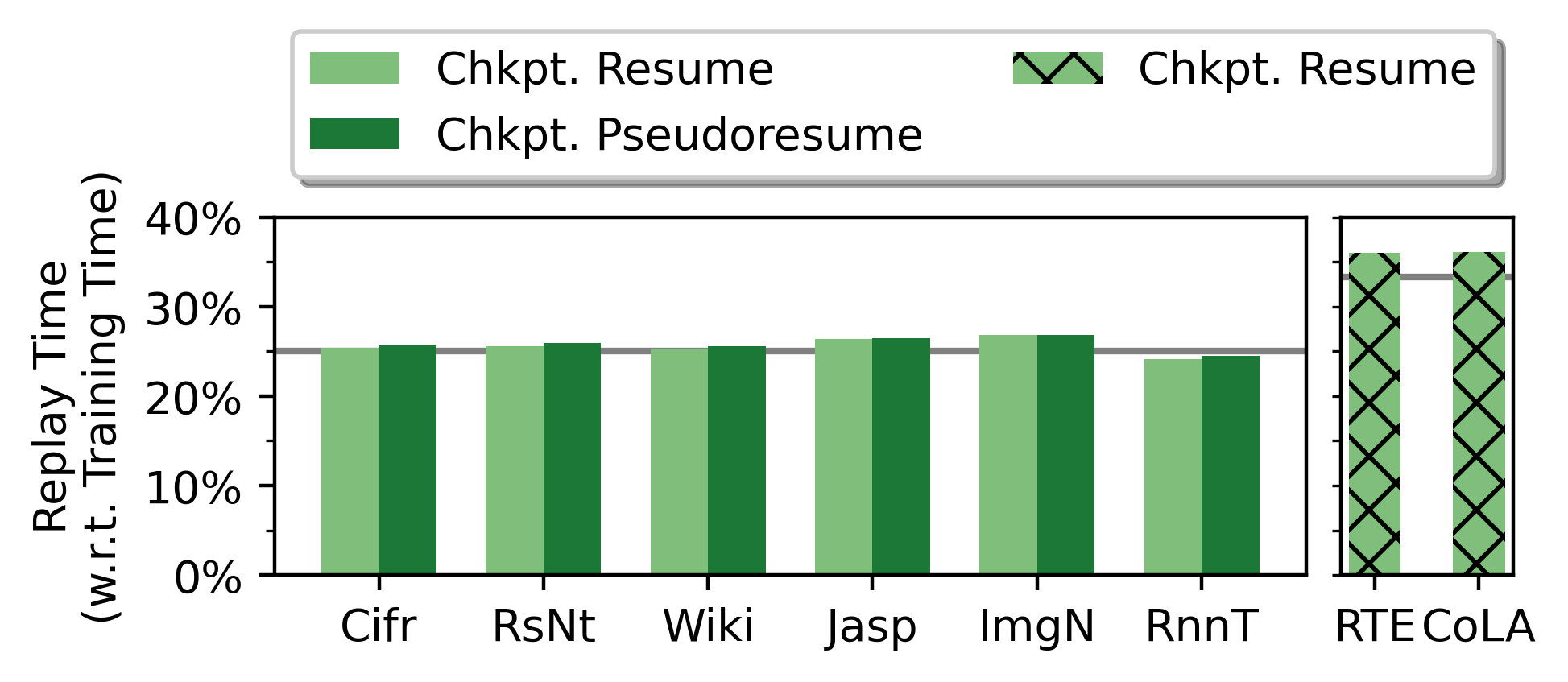}
    \caption{Parallel replay time of model training jobs (4$x$ parallelism), as fraction of a serial re-execution. 
    RTE \& CoLA only have 6 work partitions each, so parallelism on 4~GPUs leads to at best $2/6 ~= 33\%$
    replay time.
    }
    \label{fig:initialization}
\end{figure} 

\begin{figure}
    \includegraphics[width=\columnwidth]{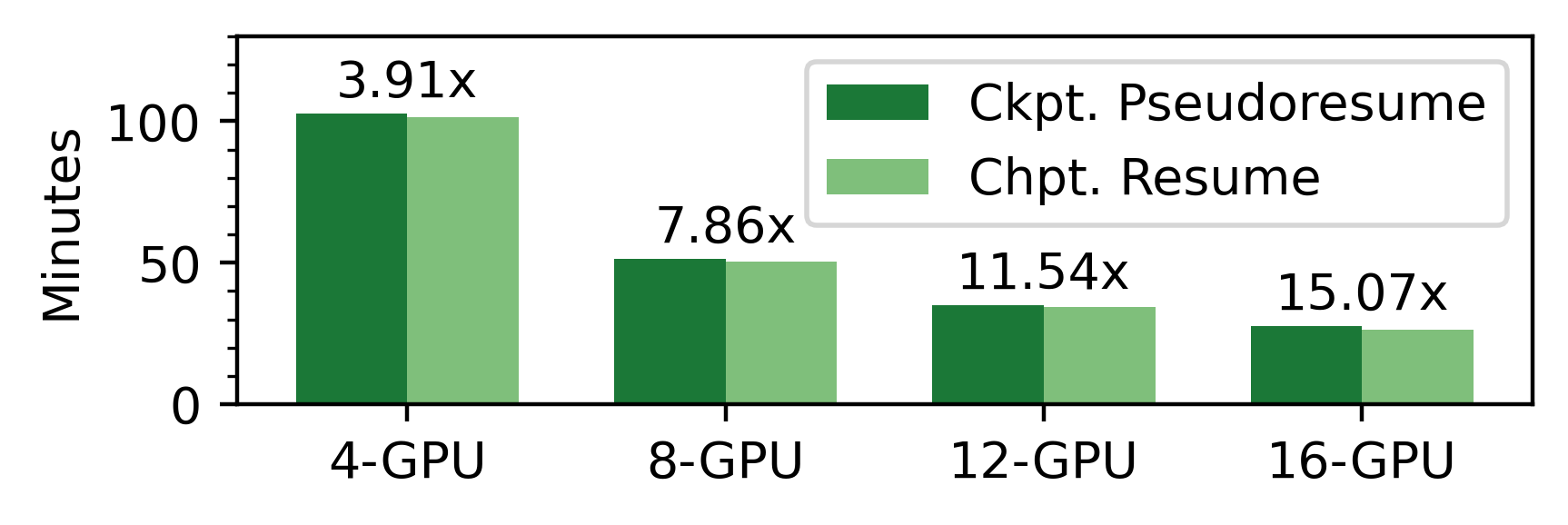}
    \caption{
    Replay time using GPUs from multiple P3.8xLarge machines, on experiment RsNt.
    The ``checkpoint resume'' speedup relative to a sequential execution is denoted by the text labels.
    }
    \label{fig:scaleout}
\end{figure}

Next, we compare the performance of parallel replay with
checkpoint resume against parallel replay with checkpoint pseudoresume (refer to Section~\ref{subsec:pseudoresume}).
An expert model developer, such as Judy, who does periodic checkpointing by-hand can ensure
that the checkpoints are complete with respect to training. Thus,
they can achieve the \textit{checkpoint resume} performance.
On the other hand, when \flor instruments training code on behalf of the developer,
it will rely on \textit{checkpoint pseudoresume}, because as we discussed earlier,
\flor cannot automatically ensure that its checkpoints are complete with respect to training,
and it assumes that the checkpoints are partial.
Our results show that, although pseudoresuming training adds initialization overhead,
this overhead is amortized through the course of parallel replay, such that there 
is a negligible difference between checkpoint resume and checkpoint pseudoresume.

In Figure~\ref{fig:initialization}, we measured parallel \replay{} performance, and observe 
it achieves near-ideal parallelism. 
Ideal parallelism is denoted by the gray horizontal line in each subplot.
Because parallel workers do not need to communicate or coordinate,
\flor \replay{} is especially well-suited for elastic and horizontally scalable cloud computing,
in which it can scale out to more GPUs at low marginal costs.
This level of parallelism is achievable because
model training replay is embarrassingly parallel given (complete or partial) checkpoints.

To assess our parallel performance, in Figure~\ref{fig:scaleout} we illustrate the near-ideal
speedup as we add 4-GPU machines. 
We choose RsNt as our experiment because it has 200 epochs to parallelize. 
The modest gap between our results and ideal here is due to load balancing limitations: 
balancing 200 epochs over 16 parallel workers results in each worker doing up to 13 epochs of work.
Consequently, the maximum achievable speedup for this workload on 16 GPUs is $\frac{200}{13}$: 15.38$\times$.

\botitem{Takeaway}: Model training replay is embarrassingly parallel given periodic checkpoints.
Parallel replay is achieved by resuming from a checkpoint. When no complete checkpoint is available,
it is possible to pseudoresume from a partial checkpoint. Although pseudoresuming adds initialization 
overhead to parallel replay, the cost is amortized over the course of replay, such that there is 
a negligible difference between checkpoint resume and partial checkpoint pseudoresume.

\section{Related Work}\label{sec:related_work}
\smallitem{ML lifecycle management.} 
The machine learning lifecycle encompasses many tasks, including
model design and development,
training, validation, deployment, inference, and monitoring~\cite{garcia2018context}.
There is a wide range of research and tooling being developed to support
these many tasks.
ML lifecycle management is especially challenging
because it involves many cycles of trial-and-error~\cite{lee2020demystifying}, 
and its dependencies are hard to scope~\cite{sculley2014machine}.
When something goes wrong, ML engineers may need 
to rollback their model to an earlier version~\cite{miao2017modelhub,vartak2016m},
inspect old versions of the training data~\cite{huang2017orpheusdb, jensen1999temporal,maddox2016decibel}, 
or audit the code that was used for training~\cite{miao2018provdb, schelter2017automatically}.
Those activities require the proper management, versioning, and provenance tracking of
data, models, code, and other context; 
existing solutions provide some 
support~\cite{baylor2017tfx, hellerstein2017ground, lee2018pretzel, liberty2020elastic, zaharia2018accelerating}.
Hindsight logging is a novel contribution in the lifecycle, and its
minimalist, low-friction interface makes it complementary to the prior work. 
\flor is designed to be compatible with any of the tools in the Python ecosystem.
In terms of training libraries, we have focused on PyTorch, but adopting
another training library involves only encoding any side-effects
in the library's API (Section~\ref{subsec:side_effect_analysis}).

\smallitem{Model Debugging.} 
There are many tools and techniques for helping users understand
the behavior of their models~\cite{google_pair, lundberg2017unified, ribeiro2016should, ribeiro2018anchors, selvaraju2017grad}, and 
for inspecting model internals~\cite{liu2016towards, rauber2016visualizing, tzeng2005opening, vartak2018mistique, olah2018building}.
These techniques only inspect the models themselves, 
and are complementary to our work here, which focuses on the
execution data generated while training the models.

The value of execution data is evidenced by widespread use of
domain-specific loggers and visualization tools for that data, including
TensorBoard~\cite{tensorboard}, MLflow Tracking~\cite{zaharia2018accelerating}, and WandB~\cite{wandb}.
Hindsight logging 
allows developers to
keep their current logging practices and tools,
but if they find they missed or forgot to log necessary execution data,
we enable them to ``query the past''. 

\smallitem{Partial Materialization.}
Inspired by classical work on materialized views~\cite{chirkova2011materialized}, a new body of work addresses partial materialization of state in ML workflows, to aid in iterative tasks like debugging. As representative examples, Columbus~\cite{zhang2016materialization} accelerates the exploration of feature selection by choosing to cache feature columns; Helix~\cite{xin2018helix} focuses on choosing to cache and reuse the outputs of black-box workflow steps; Mistique~\cite{vartak2018mistique} focuses on techniques for compressing model-related state and deciding whether to materialize or recompute. These systems introduce bespoke languages for pre-declaring what to capture prior to computation; they also provide custom query APIs to interrogate the results.
Hindsight logging is complementary: it enables post-hoc materialization in cases when it was~\emph{not} prespecified. Precisely because \flor does not dictate a new API, it is compatible with this prior work: users of these systems (or any library with pre-declared annotations) can benefit from \flor to add annotations in hindsight, and benefit from \flor{}'s efficient replay to add materialized state. At a more mechanistic level, some of the  policies and mechanisms from this work (e.g., the model compression of Mistique) could be adapted into hindsight logging context to further improve upon our results.

\smallitem{Recovery and Replay Systems.}
Our techniques are inspired by literature on both database recovery and
program replay.
Hindsight logging is a redo-only workload, and we use a 
``physiological'' approach~\cite{gray1992transaction}:
in our view, a model training script is a complete logical
log (in the WAL sense) of a model training execution, and 
occasional physical checkpoints serve solely
to speed up redo processing.
Parallel and selective redo recovery 
was studied as early as ARIES~\cite{mohan1992aries,weikum2001transactional}. 
Parallelism in those techniques is data-partitioned and recovers the most recent consistent state; we are in essence time-partitioned and recover all prior states.
In that sense our work bears a resemblance to multiversion storage schemes from  POSTGRES~\cite{stonebraker1987design} onward to more recent efforts (e.g., ~\cite{lomet2009improving, neumann2015fast}). These systems focus on storing complete physical versions, which is infeasible in our setting due to constraints
on runtime overhead.

Numerous program \record{}-\replay{} systems have been used in the past for 
less data-oriented problems. Jalangi is a system for dynamic program analysis
that automatically records the required state during 
normal processing, and enables high-fidelity
selective replay~\cite{sen2013jalangi}. 
This is achieved by identifying and storing
memory loads that may not be available at replay time, using a 
``shadow memory'' technique.
Unlike \flor{}, Jalangi replay has strict correctness guarantees.
\flor uses side-effect analysis rather than shadow memory because the
former is lighter on overhead: 
in this sense, we risk \replay{} anomalies to
reduce \record{} overhead and \replay{} latency.

Prior work on Output 
Deterministic Replay~\cite{altekar2009odr} makes a similar trade-off 
as we do. However that work pays for higher latencies to enable
reproduction of nondeterministic bugs; 
we can avoid that overhead in Python model-training scenarios because sources of non-determinism (e.g. random seeds) are
typically captured, and model-training frameworks are increasingly designed
for reproducibility.
An interesting line of work enables reverse replay with
relatively high fidelity and without overhead
by using memory dumps on a crash~\cite{cui2018rept}---this 
impressive result is made possible by the 
spatial locality of bugs in the vicinity of execution crashes; 
one complication
with model debugging is that training errors, such as over-fitting, 
may not crash the program.
We borrow the \skipblock language construct from Chasins and Bodik's
\record{}-\replay{} system for web scraping~\cite{chasins2017skip}.

\section{Conclusion}

    At every step of exploration, model developers routinely track and visualize time series data
    to assess learning.
    Most model developers log training metrics such as the loss and accuracy by default,
    but there soon arise important differences between what additional training data model developers
    log---with major implications for data management.

    In contrast to conservative logging,
    optimistic logging is an agile and lazy practice especially well-suited to 
    early and unstructured stages of exploratory model development.
    In optimistic logging, 
    model developers log training metrics such as the loss and accuracy by default,
    and defer collection of more expensive data until analysis time, when they may restore it selectively with hindsight logging.
    In the common case, or fast path, model developers get all the relevant information from the training loss, and move on.
    In exceptional cases, however, training replay may be necessary for post-hoc data restoration.
    In this paper, we documented a system of method for efficient record-replay of model training.
    Methodical hindsight logging consists of: (i) periodic checkpointing, (ii) block memoization, and (iii) checkpoint resume.
    To extend the benefits of methodical hindsight logging to novices and experts alike, we
    open source the \flor suite for hindsight logging, which includes tools for: (i) low-overhead background logging, 
    (ii) adaptable periodic checkpointing, and (iii) end-to-end instrumentation for efficient and fully automatic record-replay.
    We evaluated methodical hindsight logging to show that it achieves the goal of efficient record-replay,
    and then compared the instrumentation library provided by \flor against the methodical expert-tuned approach,
    and find that the performance is comparable.

\begin{acks}
We would like to thank the anonymous reviewers for their patient, thoughtful, and thorough feedback and guidance.
Their recommendations substantially improved the quality of the paper.
We would also like to thank Marvin Theimer and Paul Barham for helpful suggestions and discussion.
Malhar Patel, Sona Jeswani, and Karina Uppal provided valuable help with the development of earlier versions of \flor.
In addition to NSF CISE Expeditions Award CCF-1730628, 
and an NSF Graduate Research Fellowship, 
this research is supported by gifts from Alibaba, Amazon Web Services, Ant Financial, CapitalOne, Ericsson, Facebook, Futurewei, Google, Intel, Microsoft, Nvidia, Scotiabank, Splunk and VMware.
\end{acks}

\bibliographystyle{ACM-Reference-Format}
\bibliography{main}

\end{document}